\documentclass[twocolumn,aps,showpacs,superscriptaddress]{revtex4-1}
\usepackage{amsmath}
\usepackage{graphicx}
\RequirePackage[pagewise,mathlines]{lineno}

\begin{document}
\title{Photoproduction of J/$\psi$ in non-single-diffractive p+p collisions}%
\author{Z. Cao}\affiliation{State Key Laboratory of Particle Detection and Electronics, University of Science and Technology of China, Hefei 230026, China}\affiliation{Department of Modern Physics, University of Science and Technology of China, Hefei 230026, China}
\author{L. Ruan}\affiliation{Brookhaven National Laboratory, New York, USA}
\author{Z. Tang}\affiliation{State Key Laboratory of Particle Detection and Electronics, University of Science and Technology of China, Hefei 230026, China}\affiliation{Department of Modern Physics, University of Science and Technology of China, Hefei 230026, China}
\author{Z. Xu}\affiliation{Brookhaven National Laboratory, New York, USA}\affiliation{Shandong University, Jinan, China}
\author{C. Yang}\affiliation{Shandong University, Jinan, China}
\author{S. Yang}\affiliation{Brookhaven National Laboratory, New York, USA}
\author{W. Zha}\email{first@ustc.edu.cn}\affiliation{State Key Laboratory of Particle Detection and Electronics, University of Science and Technology of China, Hefei 230026, China}\affiliation{Department of Modern Physics, University of Science and Technology of China, Hefei 230026, China}
\date{\today}%
\begin{abstract}
Recently, significant enhancements of J/$\psi$ production at very low transverse momenta were observed by the ALICE and STAR collaboration in peripheral hadronic A+A collisions. The anomaly excesses point to evidence of coherent photon-nucleus interactions in violent hadronic heavy-ion collisions, which were conventionally studied only in ultra-peripheral collisions. Assuming that the coherent photoproduction is the underlying mechanism which is responsible for the excess observed in peripheral A+A collisions, its contribution in p+p collisions with nuclear overlap, i.e. non-single-diffractive collisions, is of particular interest. In this paper, we perform a calculation of exclusive J/$\psi$ photoproduction in non-single-diffractive p+p collisions at RHIC and LHC energies base on the pQCD motivated parametrization from world-wide experimental data, which could be further employed to improve the precision of phenomenal calculations for photoproduction in A+A collisions. The differential rapidity and transverse momentum distributions of J/$\psi$ from photoproduction are presented. In comparison with the J/$\psi$ production from hadronic interactions, we find that the contribution of photoproduction is negligible.
\end{abstract}
\maketitle
In ultra-relativistic heavy-ion collisions, one aims at searching for a new form of matter - the Quark-Gluon Plasma (QGP), which was predicted by the lattice Quantum Chromodynamics (QCD) calculation~\cite{PBM_QGP}, and studying its properties in laboratory. Among the probes of QGP, J/$\psi$ suppression in hadronic heavy-ion collisions with respect to that in elementary p+p collisions has been suggested as a ``smoking gun'' signature of QGP formation~\cite{MATSUI1986416} due to the color screening effect in the deconfined medium. J/$\psi$ can also be generated by the intense electromagnetic fields accompanied with the relativistic heavy ions via coherent photoproduction~\cite{UPCreview}. The coherently produced J/$\psi$ are expected to probe the nuclear gluon distribution at low Bjorken-$x$~\cite{REBYAKOVA2012647}, for which there is still considerably large uncertainty~\cite{1126-6708-2009-04-065}. Conventionally, the associate physics extracted from J/$\psi$ photoproduction and hadronic production belong to different subject field, and they are studied in ultra-peripheral collisions (UPC) and hadronic collisions individually. In UPC, only photoproduction and related physics were studied, since there is no hadronic interaction; analogously, in hadronic collisions, only hadronic production was expected, in which the coherent photoproduction was prohibited due to the ``coherent'' requirement.

Is the coherent photoproduction really prohibited in hadronic collisions, where the violent strong interactions occur? Recently, a significant excess of J/$\psi$ production at very low transverse momentum ($p_{T} < $ 0.3 GeV/c) has been observed by the ALICE collaboration in peripheral hadronic Pb+Pb collisions at forward-rapidity~\cite{LOW_ALICE}, which can not be described by the hadronic production modified by the hot and cold medium effects. STAR made the same measurements in Au+Au collisions at $\sqrt{s_{\rm{NN}}} = $ 200 GeV and U+U collisions at $\sqrt{s_{\rm{NN}}} =$ 193 GeV~\cite{1742-6596-779-1-012039}, and also observed significant enhancements at very low $p_{T}$ in peripheral collision. The observed excesses reveal characteristics of coherent photoproduction and can be quantitatively explained by the theoretical calculations with coherent photon-nucleus production mechanism~\cite{PhysRevC.93.044912,PhysRevC.97.044910,SHI2018399}, which strongly suggests the existence of coherent photoproduction in hadronic collisions. If coherent photoproduction is the underlying mechanism responsible for the observed excesses in hadronic A+A collisions, how about its contribution in hadronic p+p collision? Can we observe the excess originated from the same production mechanism in hadronic p+p collisions? If the contribution is significant, it would affect the pp baseline used for nuclear modification factor ($R_{\rm{AA}}$) of J$/\psi$, which would further bias our understanding of QGP extracted from J/$\psi$ suppression measurements. In this paper, we perform a calculation of exclusive J/$\psi$ photoproduction in non-single-diffractive (NSD) p+p collisions at RHIC and LHC energies. The differential rapidity and transverse momentum distributions of J/$\psi$ from photoproduction are presented, which will be compared to those from hadronic production.

According to the equivalent photon approximation, the photoproduction rate in p+p collisions can be factorized into two part: the photon flux, and the photon-proton cross section. The cross section can be written as:
  \begin{equation}
  \label{equation1}
  \sigma({p + p} \rightarrow {p + p} + \text{J}/\psi) = \int d\omega n(\omega)\sigma(\gamma p \rightarrow \text{J}/\psi p),
  \end{equation}
  where $\omega$ is the photon energy, $n(\omega)$ is the photon flux at energy $\omega$, and $\sigma(\gamma p \rightarrow \text{J}/\psi p)$ is the photonuclear interaction cross-section for J$/\psi$.

\begin{figure}[htbp]
\includegraphics[keepaspectratio,width=0.45\textwidth]{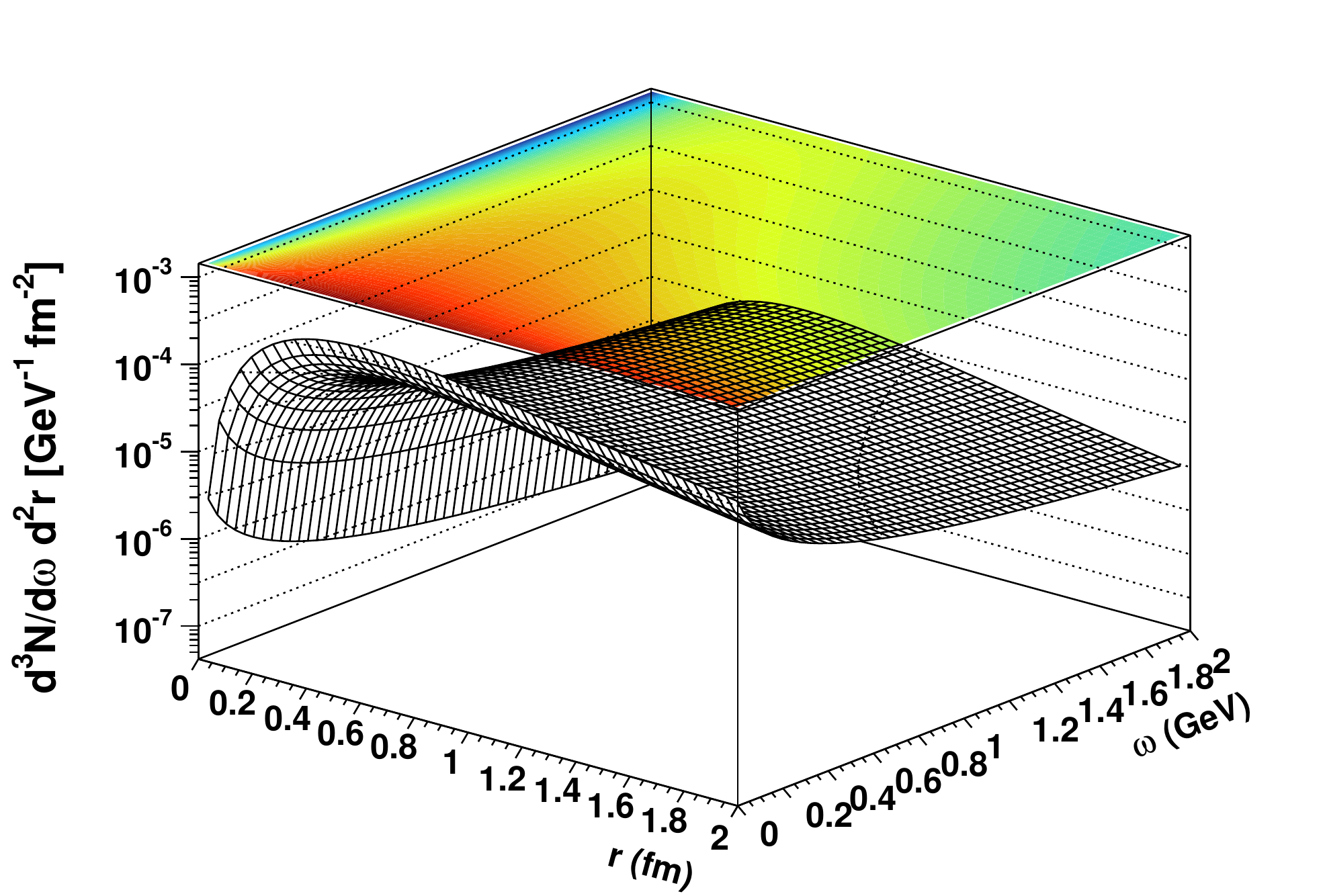}
\caption{Two-dimensional distributions of the photon flux in the distant $r$ and in the energy of photon $\omega$ for p+p collisions at $\sqrt{s}$ = 200 GeV}
\label{figure1}
\end{figure}

  The photon flux induced by proton can be modelled using the Weizs\"acker-Williams method~\cite{KRAUSS1997503}. For the point-like charge distribution, the photon flux is given by the simple formula
    \begin{equation}
    n(\omega,r) = \frac{d^{3}N}{d\omega d^{2}r} = \frac{Z^{2}\alpha}{\pi^{2}\omega r^{2}}x^{2}K_{1}^{2}(x)
         \label{equation2}
    \end{equation}
where $n(\omega,r)$ is the flux of photons with energy $\omega$ at distant $r$ from the center of proton, $\alpha$ is the electromagnetic coupling constant, $x=\omega r/\gamma$, and $\gamma$ is lorentz factor. Here, $K_{1}$ is a modified Bessel function. The point-like assumption is appropriate in UPC, however, in NSD collisions, the two colliding protons come very close to each other, the proton internal structure should be taken into account. A generic formula for any charge distribution can be written as~\cite{KRAUSS1997503}:
   \begin{equation}
  \label{equation3}
  \begin{aligned}
  & n(\omega,r) = \frac{4Z^{2}\alpha}{\omega} \bigg | \int \frac{d^{2}q_{\bot}}{(2\pi)^{2}}q_{\bot} \frac{F(q)}{q^{2}} e^{iq_{\bot} \cdot r} \bigg |^{2}
  \\
  & q = (q_{\bot},\frac{\omega}{\gamma})
  \end{aligned}
  \end{equation}
  where the form factor $F(q)$ is Fourier transform of the internal charge distribution in proton. A dipole form is employed to describe the form factor of proton, defined as:
      \begin{equation}
          F(q) = \frac{a^{4}}{(q^{2}+a^{2})^{2}}
         \label{equation4}
    \end{equation}
    where the parameter $a$ is related to the root mean square charge radius of the proton ($r_{p}$: $0.8768 \pm 69$ fm~\cite{RevModPhys.80.633}). Fig.~\ref{figure1} shows the two-dimensional distributions of the photon flux induced in p+p collisions at $\sqrt{s}$ = 200 GeV as a function of distant $r$ and energy $w$ with the dipole form factor for proton. One can observe that the photon flux drops rapidly toward $r \rightarrow 0$ inside the proton.
 \renewcommand{\floatpagefraction}{0.75}

 \renewcommand{\floatpagefraction}{0.75}
\begin{figure}[htbp]
\includegraphics[keepaspectratio,width=0.45\textwidth]{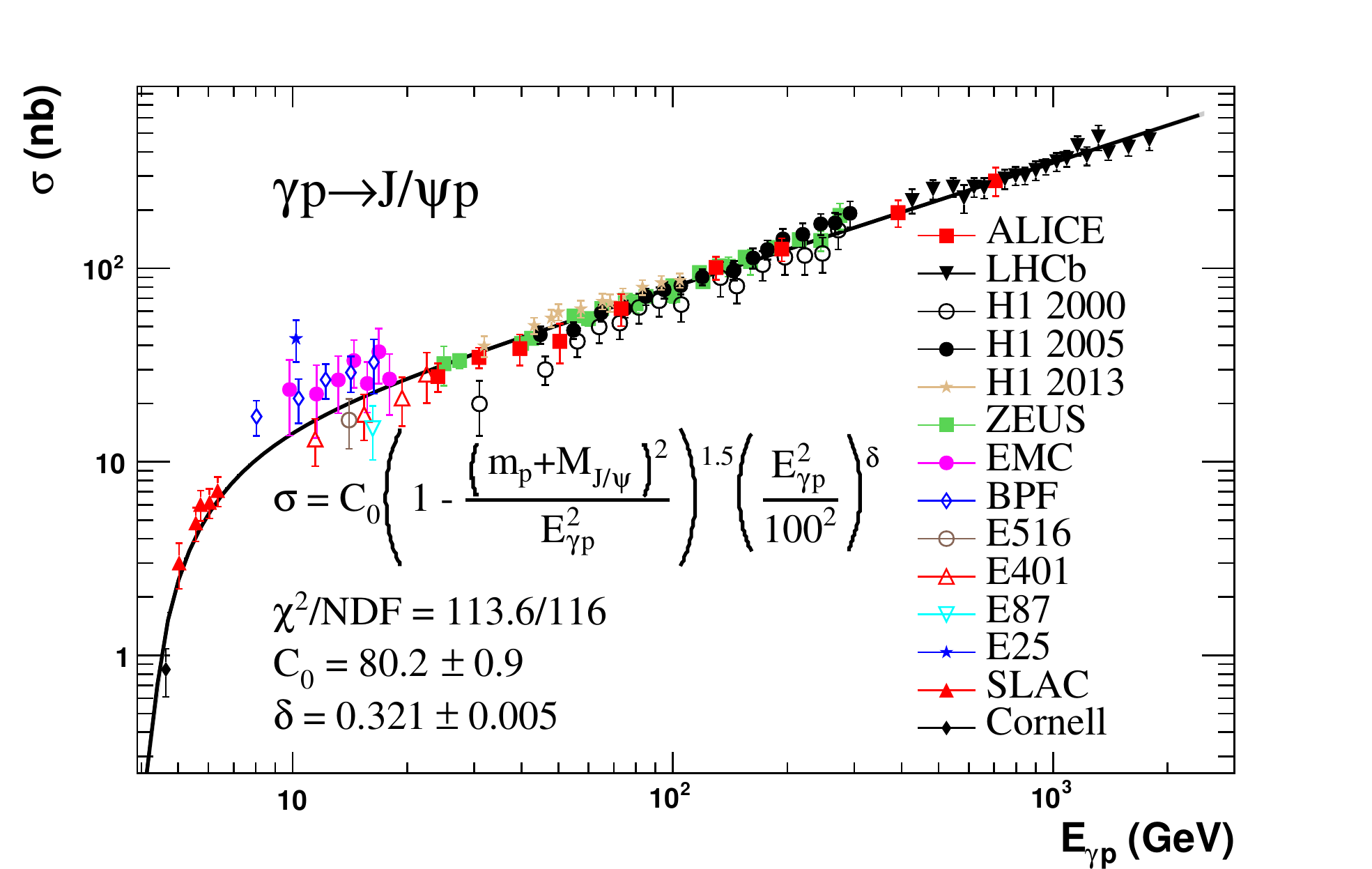}
\caption{Exclusive J/$\psi$ photoproduction cross section as a function of $E_{\gamma p}$ from world-wide experimental measurements. The black solid line with grey band on top of it represents the parametrization discussed in the text.}
\label{figure2}
\end{figure}

\begin{table}[htp]
\caption{Summary of references for world-wide data.}
\label{table1}
\centering
\begin{tabular}{c|c|c|c}
\hline
   experiment   &   $\sigma$   &   $b$     &collision system\\
\hline
ALICE  &\cite{PhysRevLett.113.232504,Acharya:2018jua}& & p-Pb/pp\\
LHCb    &\cite{Aaij:2014iea,Aaij:2018arx}&\cite{Aaij:2018arx}& pp\\
H1(2013)  &\cite{Alexa2013}&\cite{Alexa2013} & ep\\
H1(2005)  &\cite{Aktas:2005xu} &\cite{Aktas:2005xu} & ep\\
H1(2000)  &\cite{Adloff:2000vm}&\cite{Adloff:2000vm}& ep\\
ZEUS      &\cite{Chekanov:2002xi}&\cite{Chekanov:2002xi} & ep\\
EMC           &\cite{AUBERT1980267} & \cite{AUBERT1980267} & $\mu$\\
BPF           &\cite{PhysRevLett.45.682,PhysRevLett.43.187}& & $\mu$ Fe\\
E516      &\cite{PhysRevLett.52.795}&\cite{PhysRevLett.52.795} &$\gamma$ p \\
E401       &\cite{PhysRevLett.48.73}&\cite{PhysRevLett.48.73} & $\gamma$ p/d \\
E87       &\cite{PhysRevLett.34.1040,stephen-holmes} & &$\gamma$ Be\\
E25        &\cite{PhysRevLett.36.1233} &\cite{PhysRevLett.36.1233}&$\gamma$ d\\
SLAC            &\cite{PhysRevLett.35.483}  & &$\gamma$ d\\
Cornell   &\cite{PhysRevLett.35.1616} & &$\gamma$ Be\\
\hline
\end{tabular}
\end{table}

 \renewcommand{\floatpagefraction}{0.75}
\begin{figure}[htbp]
\includegraphics[keepaspectratio,width=0.45\textwidth]{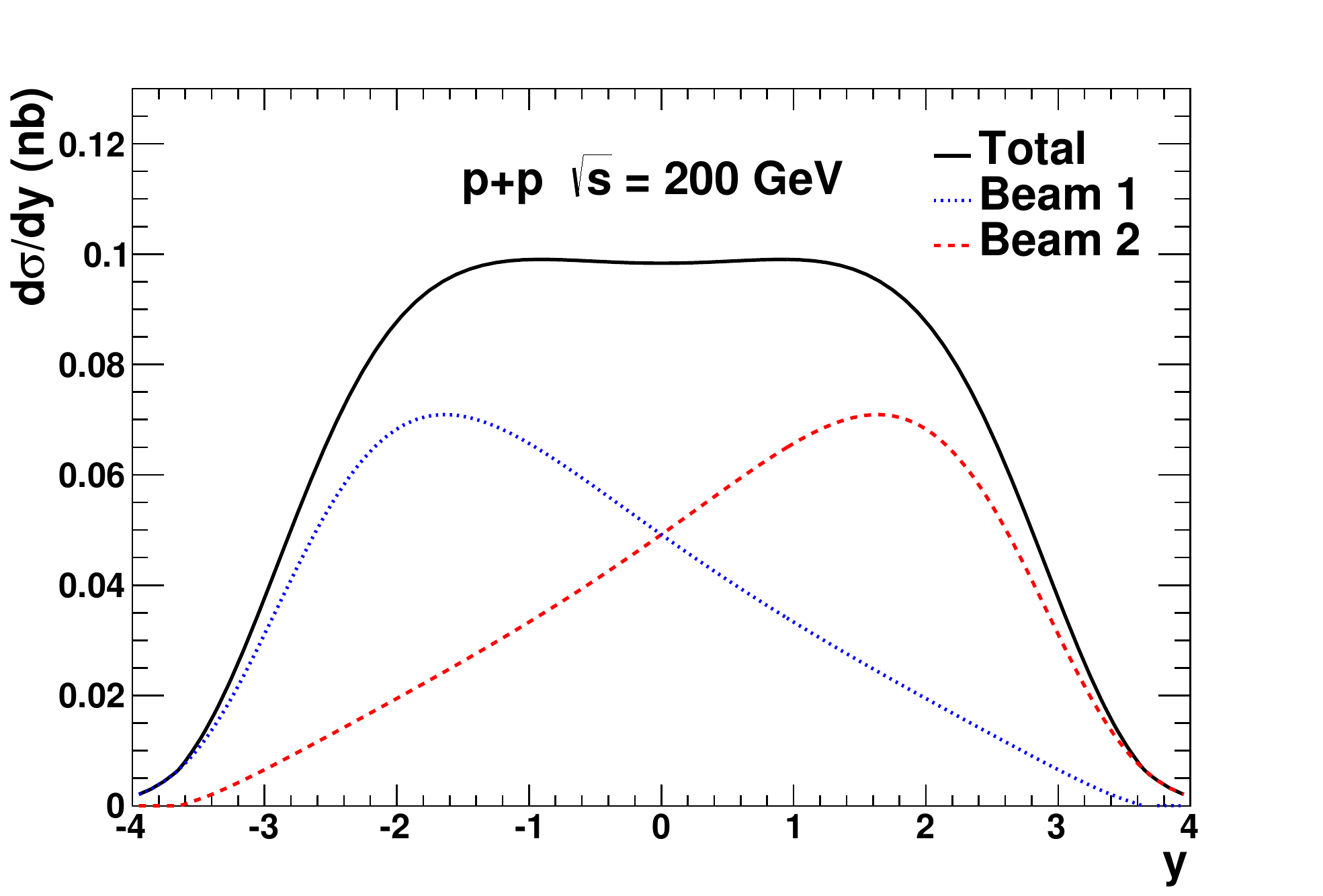}
\caption{The rapidity distribution, $d\sigma/dy$, of produced J/$\psi$ from photoproduction in NSD p+p collisions at $\sqrt{s} =$ 200 GeV. The solid line is the total production, while the dashed/dotted lines represent the individual cross section contributions from the two beams.}
\label{figure4}
\end{figure}
 \renewcommand{\floatpagefraction}{0.75}
\begin{figure*}[htbp]
\includegraphics[keepaspectratio,width=0.32\textwidth]{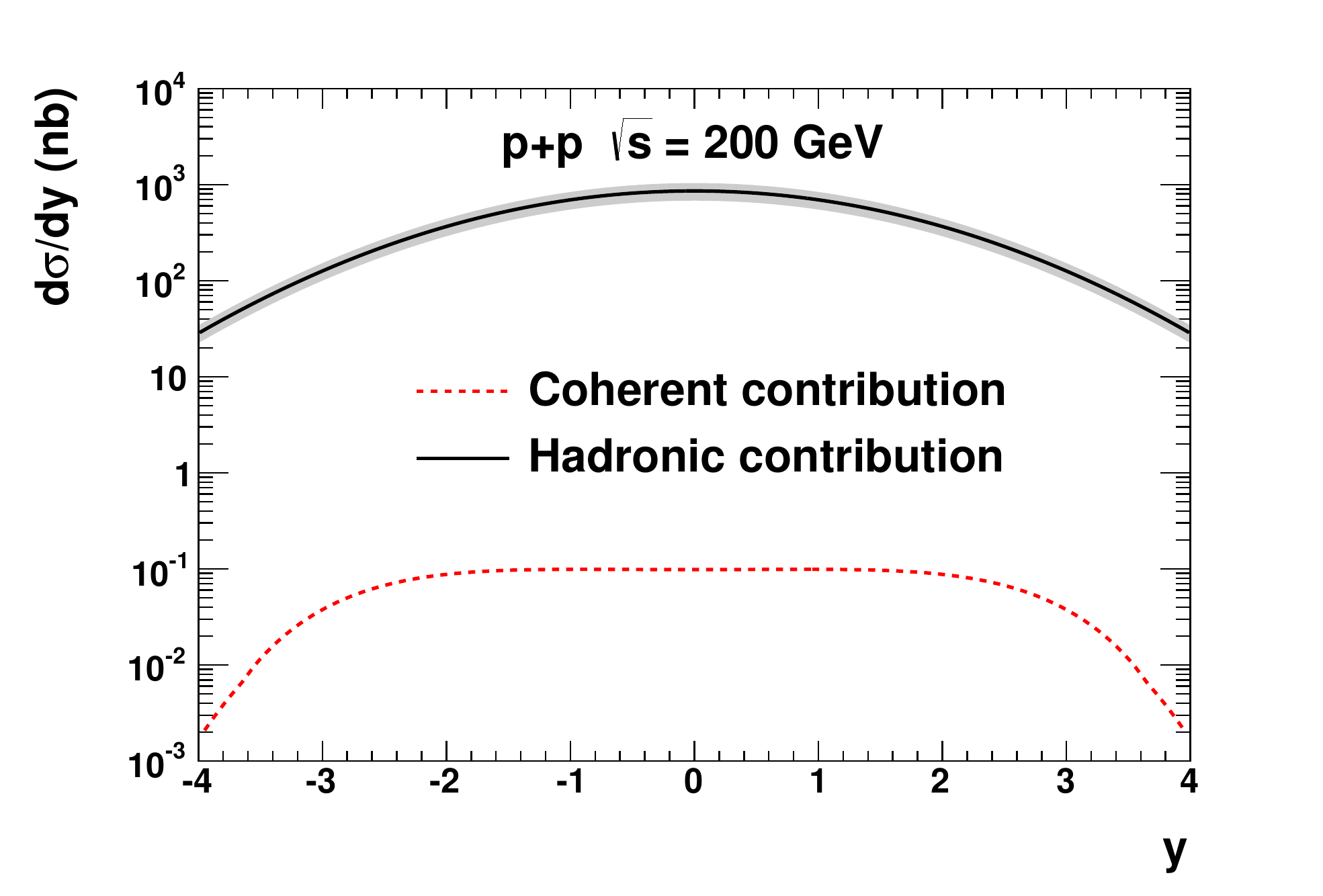}
\includegraphics[keepaspectratio,width=0.32\textwidth]{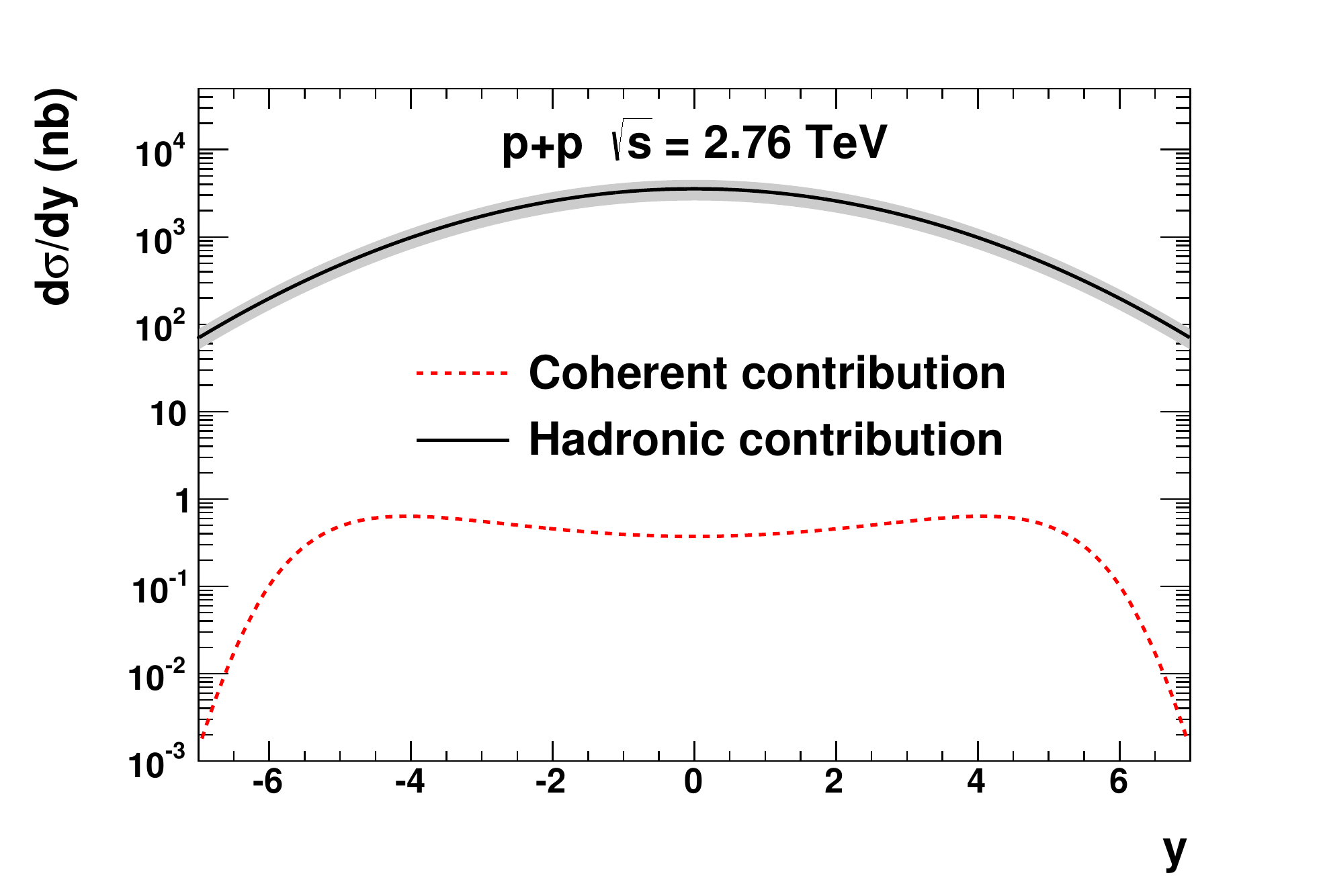}
\includegraphics[keepaspectratio,width=0.32\textwidth]{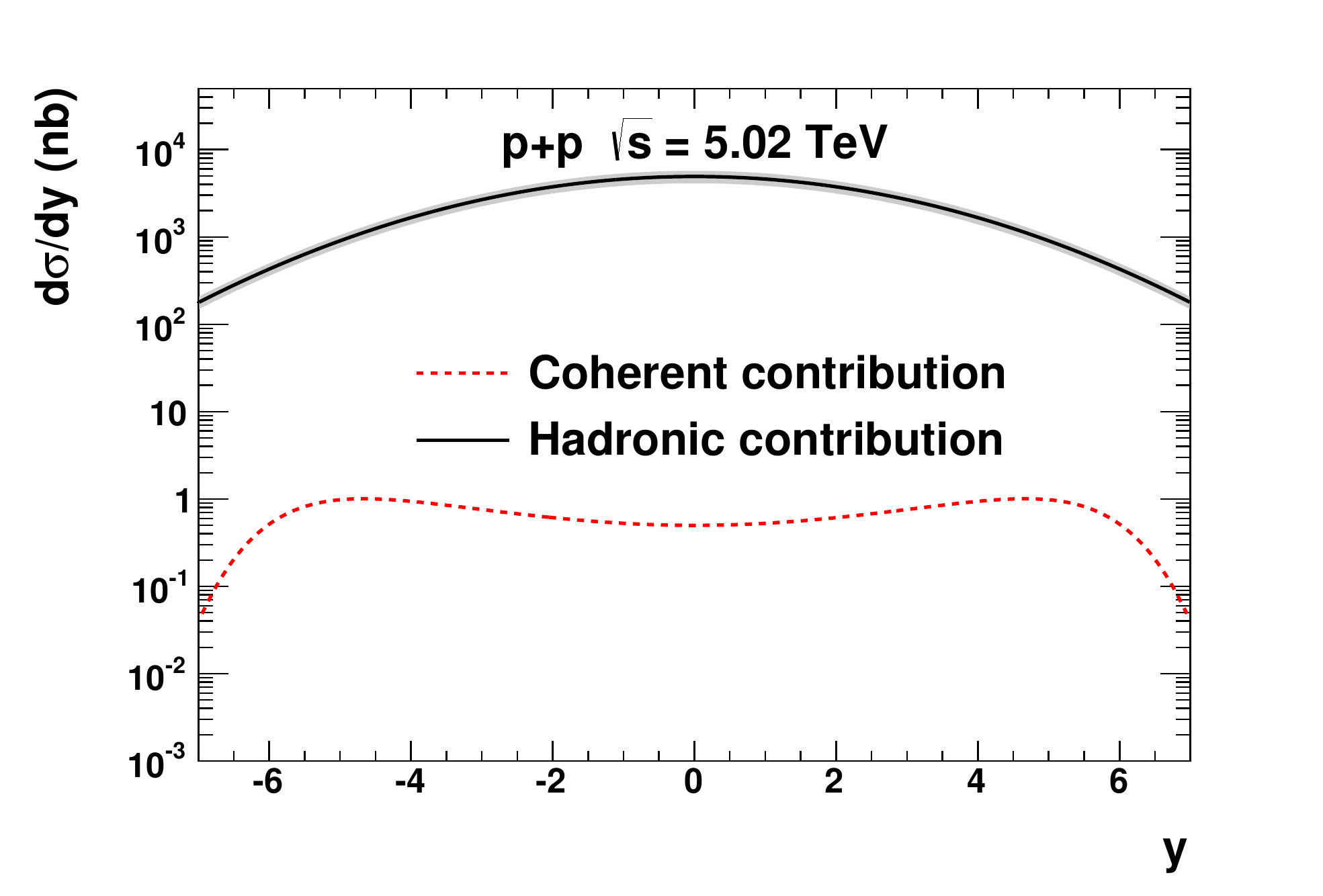}
\caption{The differential cross section of J/$\psi$ from hadronic production and photoproduction as a function of rapidity in p+p collisions at $\sqrt{s}$ = 0.2 (left panel), 2.76 (middle panel), and 5.02 TeV (right panel), respectively. The red dashed lines are predictions from photoproduction; the black solid lines with gray bands represent cross sections from hadronic production. The hadronic contributions are from parameterizations in Ref~\cite{PhysRevC.93.024919}.}
\label{figure6}
\end{figure*}

 \renewcommand{\floatpagefraction}{0.75}
\begin{figure}[htbp]
\includegraphics[keepaspectratio,width=0.45\textwidth]{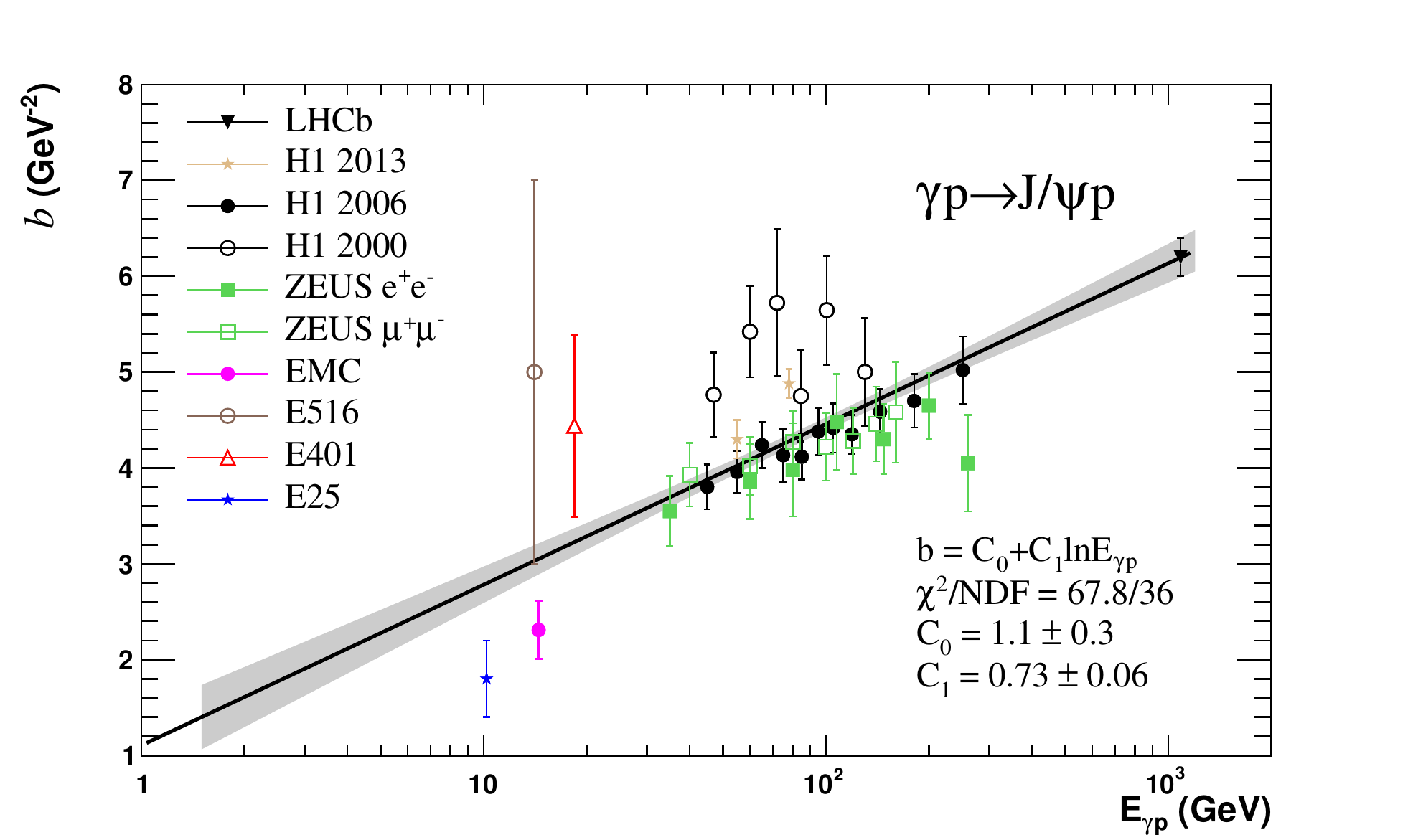}
\caption{The slope parameter ($b$) of exclusive J/$\psi$ photoproduction as a function of $E_{\gamma p}$ from world-wide experimental measurements. The black solid line with grey band represents the parametrization discussed in the text.}
\label{figure5}
\end{figure}

 \renewcommand{\floatpagefraction}{0.75}
\begin{figure*}[htbp]
\includegraphics[keepaspectratio,width=0.32\textwidth]{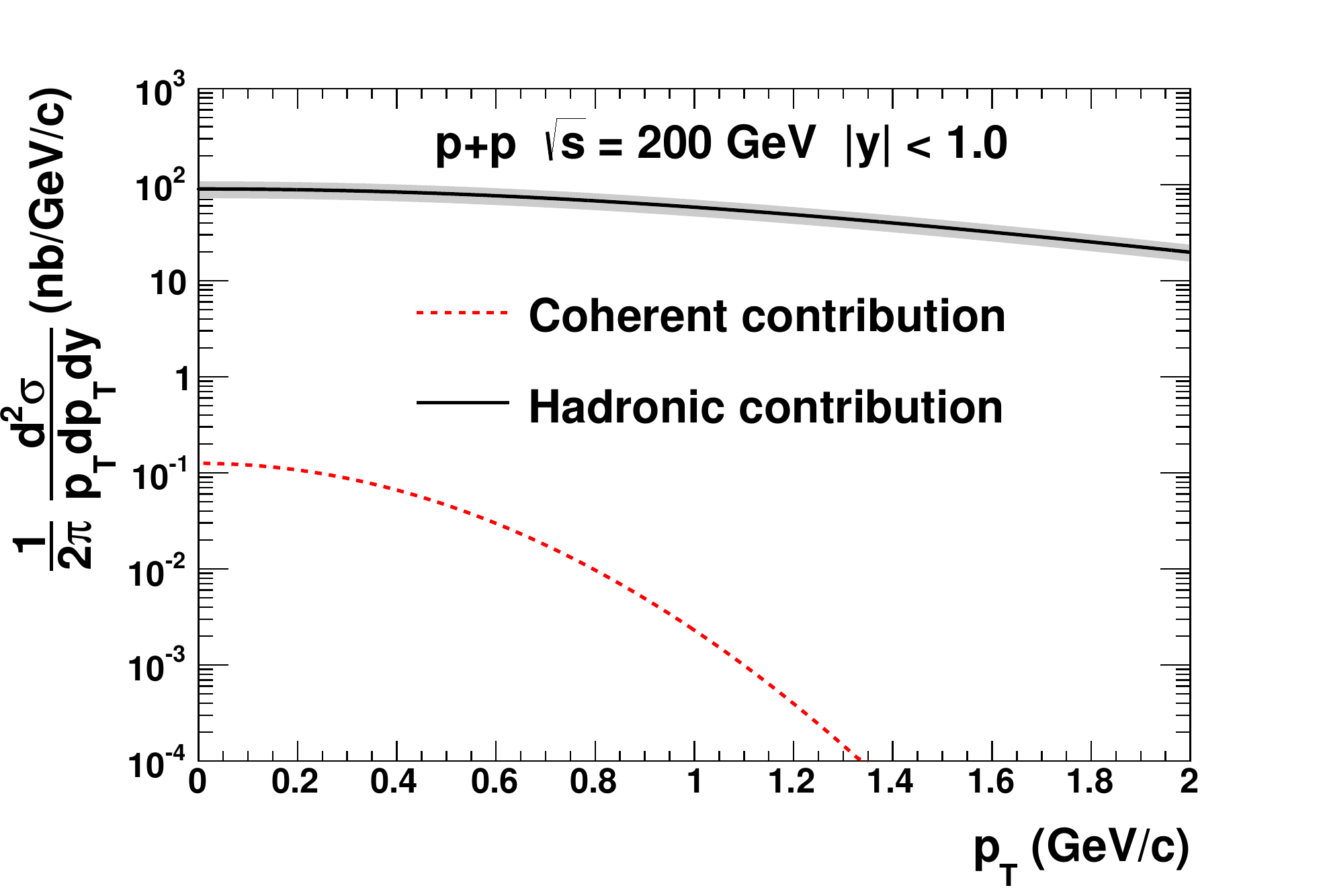}
\includegraphics[keepaspectratio,width=0.32\textwidth]{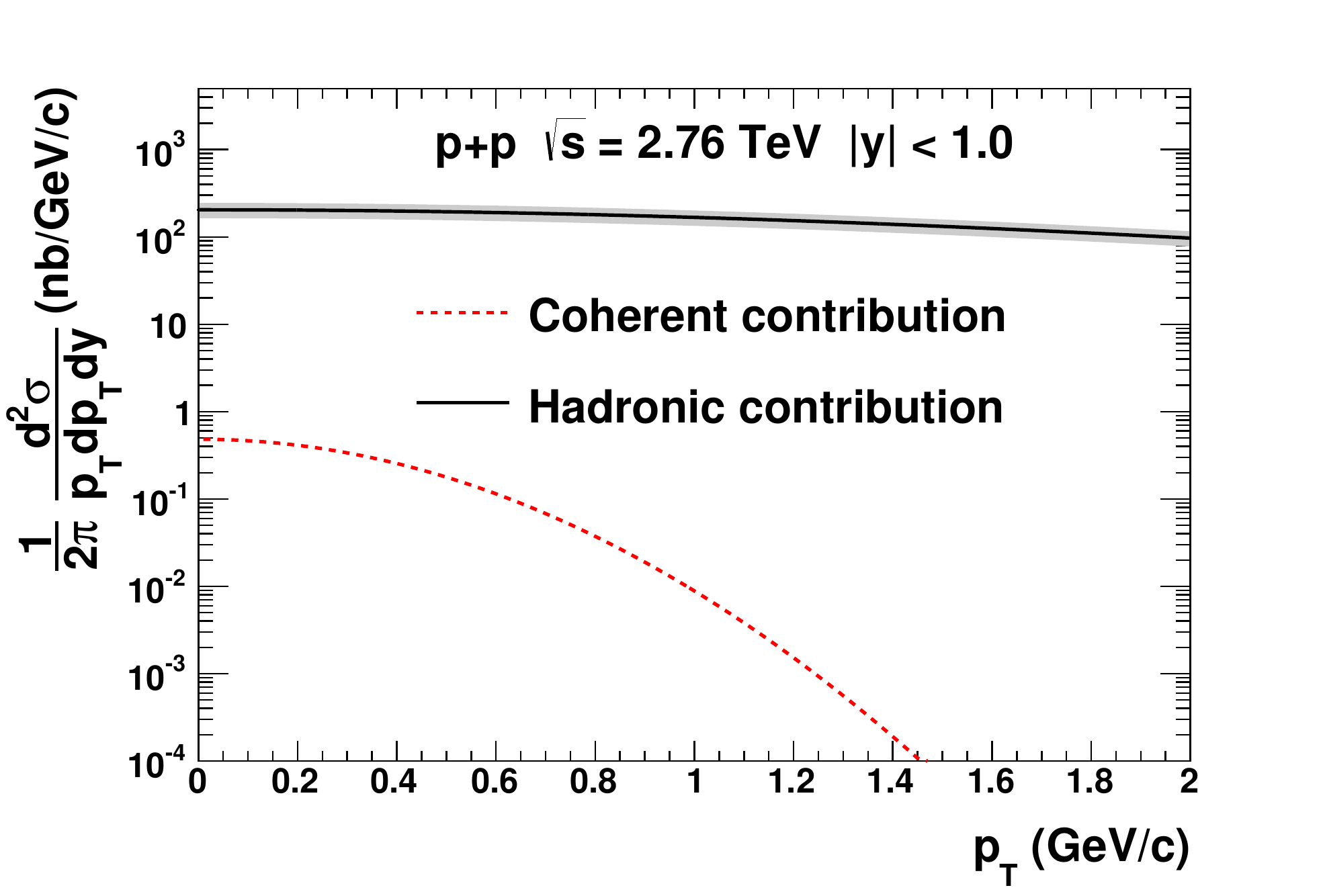}
\includegraphics[keepaspectratio,width=0.32\textwidth]{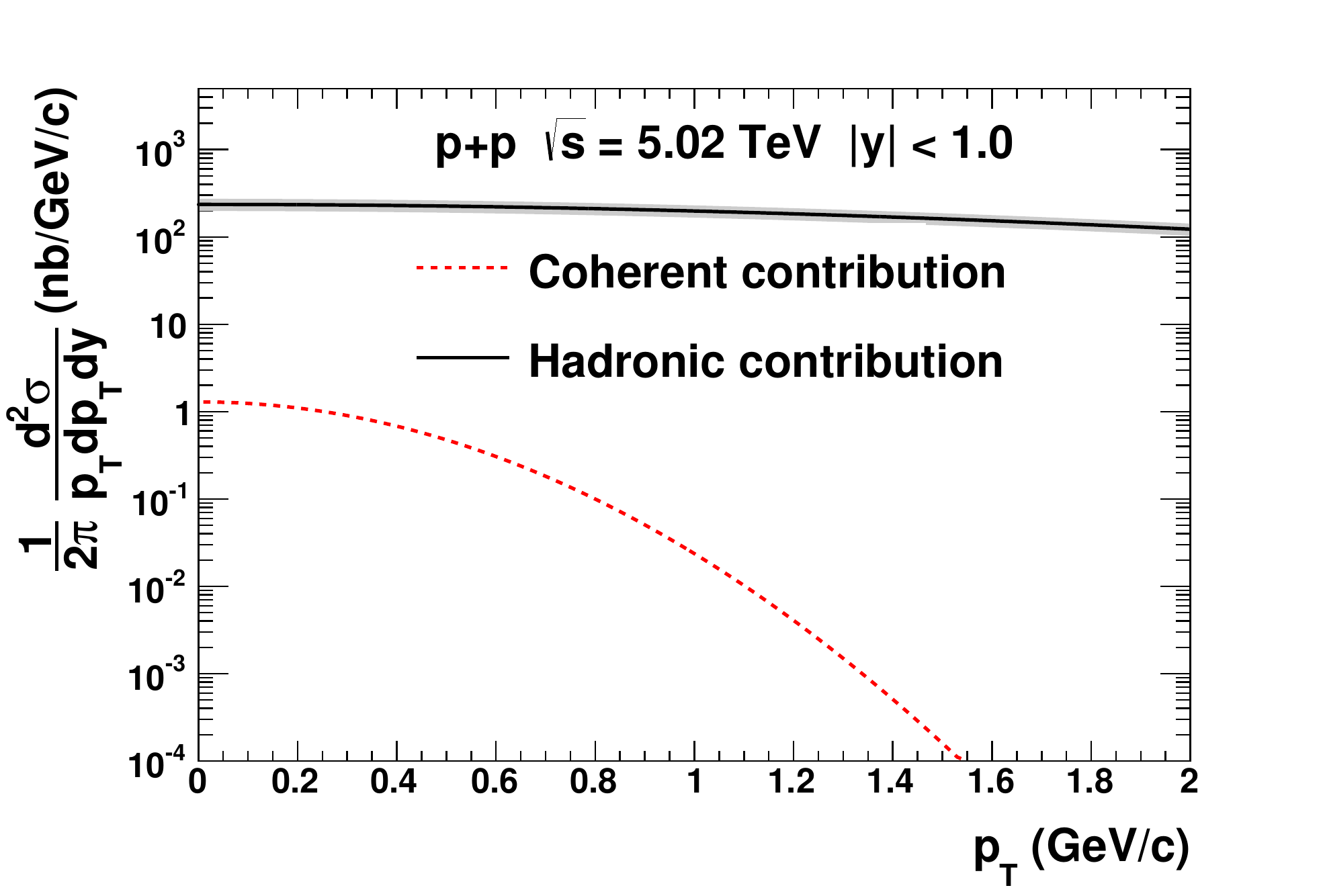}
\caption{The differential invariant cross section of J/$\psi$ from hadronic production and photoproduction as a function of transverse momentum in p+p collisions for mid-rapidity ($|y| < 1$) at $\sqrt{s}$ = 0.2 (left panel), 2.76 (middle panel), and 5.02 TeV (right panel), respectively. The red dashed lines are predictions from photoproduction; the black solid lines with gray bands represent cross sections from hadronic production. The hadronic contributions are from parameterizations in Ref~\cite{PhysRevC.93.024919}.}
\label{figure7}
\end{figure*}

The photoproduction cross sections, $\sigma (\gamma p \rightarrow \rm{J}/\psi p)$, depend on the gluon density in the proton~\cite{Ryskin1993}. At mid-rapidity, J/$\psi$ production is sensitive to gluon with $x$ down to $1.5 \times 10^{-2}$ at RHIC and $6 \times 10^{-4}$ at the LHC. However, there is still large uncertainty at such $x$ region for different PDF sets.  In this calculation, we use the world-wide experimental data on exclusive J/$\psi$ photoproduction to do the parametrization for cross section estimates. The measurements of J/$\psi$ photoproduction have been performed for more than forty years. In such a long period, different experimental techniques have been utilized and different input information was available at the time of measurements. For example, the branching ratio of $\rm{J}/\psi \rightarrow e^{+}e^{-}$ or ($\mu^{+}\mu^{-}$) have changed with time. To compare the different experimental results on an equal footing, all the measurements are updated with the latest branching fractions ($5.961 \pm 0.032\%$ for J$/\psi \rightarrow e^{+} + e^{-}$, $5.971 \pm 0.032\%$ for J$/\psi \rightarrow \mu^{+} + \mu^{-}$)~\cite{PhysRevD.98.030001}. The treated cross section as a function of $\gamma p$ center of mass energy ($E_{\gamma p}$) is shown in Fig.~\ref{figure2}. The data are fitted using the following pQCD motivated expression~\cite{STRIKMAN200572}:
\begin{equation}
\sigma \left( E_{\mathrm{\gamma p}} \right)  = \mathrm{C_0} \left( 1- \frac{\left( m_p+M_{\rm{J}/\psi} \right)^2}{E^2_{\mathrm{\gamma p}}} \right)^{1.5} \left( \frac{E^2_{\mathrm{\gamma p}}}{100^{2} \rm{GeV}^{2}} \right)^{\delta}
\label{equation5}
\end{equation}
The values of the free parameters $C_{0}$ and $\delta$ are determined from the fit, resulting in $C_{0} = 80.2 \pm 0.9$ nb and $\delta = 0.321 \pm 0.005$. The parametrization with the most complete experimental data could also be employed to improve the precision of phenomenal calculations for photoproduction in A+A collisions such as ~\cite{PhysRevC.93.044912,PhysRevC.97.044910,SHI2018399,Zha:2018ytv,PhysRevC.60.014903}.  As shown in the figure, the parameterization describes the experimental measurements very well with $\chi^{2}/NDF = 113.6/116$. The references of the data are summarized in Table~\ref{table1}.

To efficiently relate the NSD cross section to its corresponding region in impact parameter space, a purely geometrical picture is employed in the calculation:
\begin{equation}
\sigma_{\rm{NSD}} = \int_{0}^{b_{\rm{max}}}2 \pi b db = \pi b_{max}^{2} \\
\label{equation6}
\end{equation}
In this paper, we perform calculations in NSD p+p collisions at $\sqrt{s} =$ 0.2, 2.76 and 5.02 TeV, the corresponding NSD cross sections are 30, 50,and 56 mb~\cite{Abelev2013}, respectively.

With the convolution of equivalent photon spectra and elementary $\gamma p \rightarrow \rm{J}/\psi p$ cross section, the probability to produce a J/$\psi$ with rapidity $y$ for a collision at impact parameter $b$ can be given by:
\begin{equation}
\frac{dP(y,b)}{dy} = \omega N(\omega, b) \sigma_{\gamma p \rightarrow \rm{J}/\psi p}(E_{\gamma p})
\label{equation6}
\end{equation}
where $N(\omega,b)$ is the effective photon flux with impact parameter $b$ at photon energy $\omega$. The effective photon flux, $N(\omega, b)$, can be expressed through the photon flux induced by one proton and effective strength for the photon with the second proton:
\begin{equation}
N(\omega,b) = \int n(\omega, r) \frac{\theta(r_{p}-(|\vec{r} - \vec{b}|))}{\pi r_{p}^{2}}d^{2}r
\label{equation7}
\end{equation}
where $b$ is the impact parameter between the two colliding protons, $r$ is the distant from the proton which emits the photon, and the extra $\theta(r_{p}-(|\vec{r} - \vec{b}|))$ ensures collision when the photon hits the proton.
The photon energy, $\omega$, can be determined from the rapidity of J/$\psi$, $y$:
\begin{equation}
\omega = \frac{1}{2} M_{\rm{J}/\psi} e^{y}
\label{equation8}
\end{equation}
One complication is that either beam particle is equally likely to produce the photon; the cross sections for these two possibilities from two beam directions are added:
\begin{equation}
\frac{d\sigma}{dy} = \int^{b_{\rm{max}}}_{0}  (\frac{dP(y,b)}{dy} + \frac{dP(-y,b)}{dy}) 2 \pi b d^{2}b
\label{equation9}
\end{equation}
where $b_{\rm{max}}$ can be obtained from Eq.~\ref{equation6}. Figure~\ref{figure4} shows the calculated rapidity distribution, $d\sigma/dy$, of produced J/$\psi$ from photoproduction in NSD p+p collisions at $\sqrt{s} =$ 200 GeV. The solid line is the total production, while the dashed/dotted lines represent the individual cross section contributions from the two beam protons. The rapidity distribution is determined by the evolution of photon flux with photon energy $\omega$ and elementary $\gamma p$ cross section with center of mass energy $E_{\gamma p}$ at different rapidities.

Figure~\ref{figure6} shows the differential cross section of J/$\psi$ from hadronic production and photoproduction as a function of rapidity in p+p collisions at $\sqrt{s}$ = 0.2 (left panel), 2.76 (middle panel), and 5.02 TeV (right panel), respectively. The red dashed lines are predictions from photoproduction. The calculations are performed in NSD collisions, in which there exists violent strong interactions to produce J/$\psi$. The black solid lines with gray bands in the plots represent J/$\psi$ cross sections from hadronic production. The hadronic contributions are extracted from parameterizations using the world-wide experimental data, as described in Ref~\cite{PhysRevC.93.024919}. The rapidity distributions from photoproduction are different in different collision energies due to evolution of the two component structures (shown in Fig.~\ref{figure4}) from the two beam directions. In comparison with the contribution from hadronic interactions, the yield from photoproduction is several orders of magnitude lower, which makes it very difficult to detect the possible J/$\psi$ photoproduction in NSD p+p collisions.

 Could we observe an excess of J/$\psi$ at low $p_{T}$ in NSD p+p collisions similar to those in peripheral A+A collisions? Although the total cross section from photoproduction is very small in comparison to that from hadronic contribution, the J/$\psi$ from photoproduction is mainly produced at low $p_{T}$, which may gain certain significance. The J/$\psi$ $p_{T}$ from photoproduction in p+p collisions depends on the $p_{T}$ of the photon and the $p_{T}$ acquired when the vector meson is created;  and the latter is dominant. The $p_{T}$ of the photon induced by proton can be given by the equivalent photon approximation \cite{KRAUSS1997503}:
  \begin{equation}
  \label{equation14}
  \frac{d^{2}N_{\gamma}}{d^{2}\vec{k}_{\gamma\bot}} = K_{0}\frac{F_{\gamma}^{2}(\vec{k}_{\gamma})\vec{k}^{2}_{\gamma\bot}}{(\vec{k}_{\gamma\bot}^{2}+\omega^{2}_{\gamma}/\gamma_{c}^{2})^{2}}
  \end{equation}
where $F_{\gamma}(\vec{k}_{\gamma})$ is the form factor of proton used previously, $K_{0}$ is the dimensionless normalization factor, and $\vec{k}_{\gamma\bot}$ is the transverse momentum of the photon. The $p_{T}$ from the vector meson production can be estimated from the world-wide $p_{T}$ differential cross section measurements. The measured $p_{T}$ distributions can be phenomenally described by:
  \begin{equation}
  \label{equation15}
\frac{d\sigma}{dp_{T}} = N_{0} p_{T} e^{-bp_{T}^{2}}
  \end{equation}
where $N_{0}$ is the dimensionless normalization factor, $b$ is the slope parameter depending on the $\gamma p$ center of mass energy ($E_{\gamma p}$). Fig.~\ref{figure5} shows the slope parameter ($b$) of exclusive J/$\psi$ photoproduction as a function of $E_{\gamma p}$ from world-wide experimental measurements. The references of the data are summarized in Table~\ref{table1}. The black solid line with grey band represents the parametrization discussed in the following. Within the framework of Regge phenomenology~\cite{Collins:1977jy}, the slope parameter $b$ should increase logarithmically with $E_{\gamma p}$. Therefore the data are fitted using the following expression:
  \begin{equation}
  \label{equation16}
b = C_{0} + C_{1}\rm{ln}E_{\gamma p}
  \end{equation}
where $C_{0}$ and $C_{1}$ are free parameters. The corresponding $E_{\gamma p}$ is uniquely determined by the rapidity of J/$\psi$ and the collision energy of pp system. As demonstrated in the figure, the expression describes the data reasonably well. We assume that the photon $p_{T}$ and that from vector meson production are randomly oriented.

Figure~\ref{figure7} shows the differential invariant cross section of J/$\psi$ from hadronic production and photoproduction as a function of transverse momentum in p+p collisions for mid-rapidity ($|y| < 1$) at $\sqrt{s}$ = 0.2 (left panel), 2.76 (middle panel), and 5.02 TeV (right panel), respectively. The red dashed lines are predictions from photoproduction; the black solid lines with gray bands represent cross sections from hadronic production. The hadronic contributions are extracted from parameterizations using the world-wide experimental data, as described in Ref~\cite{PhysRevC.93.024919}. As depicted in the figure, even at $p_{T} \rightarrow 0$, the photoproduction contribution is several orders of magnitude smaller than that from hadronic interactions, which means that the excess originated from photoproduction at low $p_{T}$ is not visible in NSD p+p collisons at RHIC and LHC energies. However, this is a good news for the current $R_{\rm{AA}}$ measurements for very low $p_{T}$ in A+A collisions at RHIC and LHIC, since the used pp baseline for such $p_{T}$ region are from extrapolations utilizing the relative high $p_{T}$ measurements, which ignores the possible excess originated from photoproduction.

In summary, we perform a calculation of exclusive J/$\psi$ photoproduction in NSD p+p collisions at RHIC and LHC energies. The differential rapidity and transverse momentum distributions of J/$\psi$ from photoproduction are presented. In comparison with the J/$\psi$ production from hadronic interactions, the contribution of photoproduction is negligible, which suggests that, in contrast with the case in peripheral A+A collisions, the excess of J/$\psi$ yield from photoproduction is not visible in NSD p+p collisions.

We thank Dr. Spencer Klein and Prof. Pengfei Zhuang for useful discussions. This work was funded by the National Natural Science Foundation of China under Grant Nos. 11775213, 11505180 and 11675168, the U.S. DOE Office of Science under contract No. DE-SC0012704, and MOST under Grant No. 2014CB845400.

\nocite{*}
\bibliographystyle{aipnum4-1}
\bibliography{aps}
\end{document}